\documentclass[journal]{IEEEtran}

\ifCLASSINFOpdf
\else
   \usepackage[dvips]{graphicx}
\fi
\usepackage{url}

\usepackage{graphicx}
\usepackage{algorithm}
\usepackage{algpseudocode}
\usepackage{mathtools,amsmath,amsfonts,amssymb,amsthm}
\usepackage{hyperref}
\usepackage{cleveref}

\algrenewcommand\algorithmicindent{1.0em}

\usepackage[dvipsnames]{xcolor}
\usepackage{tikz}
\usepackage{pgfplots}
\pgfplotsset{compat=1.18}
\pgfplotsset{clean/.style={axis lines*=left,
        axis on top=true,
        axis x line shift=0.0em,
        axis y line shift=0.5em,
        every tick/.style={black, thick},
        axis line style = ultra thick,
        tick align=outside,
        clip=false,
        major tick length=4pt}}
\usepackage{cbcolors}

\begin{document}

\title{A divide and conquer strategy for multinomial particle filter resampling}

\author{Andrey A. Popov
\thanks{This work was supported in part by startup funds from the University of Hawai'i.}
\thanks{A. A. Popov is with the Department of Information and Computer Sciences, University of Hawai'i at Manoa, Honolulu, HI 96822 USA (e-mail: apopov@hawaii.edu).}}

\markboth{Preprint}
{Popov \MakeLowercase{\textit{et al.}}: A divide and conquer strategy for multinomial particle filter resampling}
\maketitle

\begin{abstract}
This work provides a new multinomial resampling procedure for particle filter resampling, focused on the case where the number of samples required is less than or equal to the size of the underlying discrete distribution. This setting is common in  ensemble mixture model filters such as the Gaussian mixture filter. We show superiority of our approach with respect two of the best known multinomial sampling procedures both through a computational complexity analysis and through a numerical experiment.
\end{abstract}


\IEEEpeerreviewmaketitle

\section{Introduction}

\IEEEPARstart{I}{t} is known, but not widely, that monomial linear time resampling from a discrete distribution is possible.
This work aims to correct and generalize the linear time algorithm used for multinomial particle filter resampling to make use of binary search allowing for superior performance in the setting when the number of samples taken is less than the size of the discrete distribution from which the samples originate.

We now define the problem of interest:
assume that there exists a discrete distribution defined by the probability mass function,
\begin{equation}\label{eq:pmf}
    p(s_j) = w_j, \quad j = 1, \dots, M,
\end{equation}
where each $s_j$ is some value of interest, $w_j$ is the corresponding weight with a total of $M$ weights.
The goal of this work is to sample $N$ values from this distribution with corresponding weights $1/N$ using a multinomial sampler. This means that the samples are generated using, 
\begin{equation}\label{eq:cmf}
    r_i = P^{-1}(u_i), \quad i = 1, \dots N,
\end{equation}
where $r_i$ are the samples, $u$ are uniformly distributed random variables, and $P^{-1}$ is the inverse of the cumulative mass function of~\cref{eq:pmf}, defined as,
\begin{equation}
\begin{gathered}
    P^{-1}(u) = \begin{cases}
        s_1 & u \leq W_1\\
        s_i & W_{i-1} < u \leq W_i
    \end{cases},\\ \quad W_i = \sum_{j=1}^i w_j,\quad i = 1, \dots, N,
\end{gathered}
\end{equation}
where the $W_i$ are the cumulative weights, which can be computed at the same time as the normalization of the $w_i$ with little additional cost for all particle filters.
The focus of this work is to provide a more efficient algorithm for using this `multinomial sampling procedure' in~\cref{eq:cmf}. 
Without loss of generality this paper assumes that the values of interest are sequential indices, meaning that,
\begin{equation}
    s_i = i, \quad i = 1, \dots, N,
\end{equation}
as it is trivial to convert between indices and the actual values of interest.

In the classical bootstrap particle filter~\cite{zanettisurvey} (and related particle filters) a collection of $M$ weighted particles is resampled into an unweighted collection of $N$ particles, such that, typically, $M=N$.
In this typical case, the best known algorithms~\cite{li2015resampling,carpenter1999improved} are linear in $M$. 
This work is motivated by new particle filtering algorithms such as the ensemble Gaussian mixture filter (EnGMF)~\cite{anderson1999monte, yun2019sequential, popov2024adaptive}.
In the EnGMF, the posterior is approximated by a Gaussian mixture of size $N$ times $P$ where $P$ is the Gaussian mixture that approximates the measurement likelihood. This means that when attempting to sample from said Gaussian mixture, the Gaussian component is sampled from a discrete distribution made up of $M = N\cdot P$ weights. If several measurements where incorporated in one step, this could potentially mean that $M$ is significantly larger than $N$. 
If the number of samples is kept constant at $N$, then an efficient resampling algorithm that can handle resampling $N \ll M$ particles needs to be created.

The best known algorithm for the case that $M=N$ is that due to~\cite{carpenter1999improved}, which we trivially intuit is linear $M$, but is suboptimal in the case that $M \gg N$.
This work generalizes this algorithm to the case that $M \gg N$ through a divide and conquer approach that also leverages, and takes advantage of binary-search based algorithms~\cite{li2015resampling}.
This work is organized as follows: first we introduce a few preliminary algorithms in~\cref{sec:preliminaries}, next we introduce the new divide and conquer approach in~\cref{sec:new-algorithm}, then we have a small numerical experiment with the EnGMF in~\cref{eq:numerical-experiment}, and finally conclude with a few remarks in~\cref{sec:conclusion}.

\section{Preliminaries}
\label{sec:preliminaries}
The first preliminary algorithm that we discuss is that of generating an array of sorted uniformly distributed random variables.
\begin{algorithm}
\begin{algorithmic}
\Require No. samples $N$
\Ensure $\{U_i\}_{i=1}^N = \texttt{SortedUniform}(N)$
\For{$i = 1, \dots N+1$}
\State $z_i \gets -\log(\texttt{rand}[0, 1])$
\EndFor
\For{$i = 1, \dots N+1$}
\State $Z_i \gets \sum_{j=1}^i z_j$
\EndFor
\For{$i = 1, \dots N$}
\State $U_i \gets Z_i/Z_{i+1}$
\EndFor
\end{algorithmic}
\caption{Generate sorted array of uniformly distributed random numbers through an exponential method}\label{alg:sorted-uniform}.
\end{algorithm}
We make the assumption that we have access to a function that provides uniformly distributed random samples in the interval $[0, 1]$, called $\texttt{rand}[0,1]$. The algorithm that we make use of is based on the work appearing in~\cite{lurie1972machine,rabinowitz1974comparison,gerontidis1982monte}. The algorithm takes advantage of the fact that the distance between any two uniform random variables is exponentially distributed. A pseudocode is given in~\cref{alg:sorted-uniform}.
This algorithm notably has a time complexity of $O(N)$ which is cheaper than the complexity of generate-then-sort of $O(N\log_2 N)$ (though in theory a Radix sort can equal the time complexity~\cite{terdiman2000radix}).

The second algorithm that we require is a modified binary search~\cite{knuth1998art} which find the index of the upper bound of a value given a sorted list of interval bounds. A pseudo-code of this algorithm is given in~\cref{alg:binary-search}.
\begin{algorithm}
\begin{algorithmic}
\Require Sorted list $\{Z_j\}_{j=1}^M$, No. samples $N$
\Ensure $b = \texttt{BinarySearch}(\{Z_j\}_{j=1}^M, X)$
\State $a \gets 1$
\State $b \gets M$
\State $m \gets \lfloor(a+b)/2\rfloor$
\While{$a \not= b$}
    \If{$Z_m < X$}
        \State $a \gets m + 1$
    \Else
        \State $b \gets m$
    \EndIf
    \State $m \gets \lfloor(a + b)/2\rfloor$
\EndWhile

\end{algorithmic}
\caption{Modified Binary Search}\label{alg:binary-search}
\end{algorithm}

We now move on to describing two classical multinomial sampling algorithms.
The first algorithm is based on the binary search and does not take any advantage of the fact that we are attempting to create $N$ samples.
\begin{algorithm}
\begin{algorithmic}
\Require Cumulative weights $\{W_j\}_{j=1}^M$, No. samples $N$
\Ensure $\texttt{BinarySampler}(\{W_j\}_{j=1}^M, N) = \{r_i\}_{i=1}^N$ samples
\For{$i = 1, \dots N$}
\State $U_i \gets \texttt{rand}[0, 1]$
\State $r_i \gets \texttt{BinarySearch}(\{W_j\}_{j=1}^M, U_i)$
\EndFor
\end{algorithmic}
\caption{Binary search sampler}
\end{algorithm}
The binary search sampler performs a binary search for each sample, meaning that its computational complexity is $O(N\log_2 M)$.

The Carpenter-Clifford-Fearnhead sampler~\cite{carpenter1999improved}, which we abbreviate CCF, assumes that we wish to generate $N$ samples from the target distribution and therefore takes advantage of this fact by generating a sorted array of uniformly distributed random variables. The algorithm proceeds as follows: i) the cumulative distribution is computed from the weights, ii) an array of sorted uniformly-distributed random variables is computed using any algorithm such as~\cref{alg:sorted-uniform}, iii) a linear search is conducted to determine the index of each sample corresponding to the value of the cumulative distribution.
An algorithmic description of this algorithm appears in~\cref{alg:Carpenter}.
\begin{algorithm}
\begin{algorithmic}
\Require Cumulative weights $\{W_j\}_{j=1}^M$, No. samples $N$
\Ensure $\texttt{CCFSampler}(\{W_j\}_{j=1}^M, N) = \{r_i\}_{i=1}^N$ samples
\State $\{U_i\}_{i=1}^N \gets \texttt{SortedUniform}(N)$
\State $j \gets 1$
\For{$i = 1,\dots,N$}
    \While{$W_j < U_i$}
        \State{$j \gets j + 1$}
    \EndWhile
    \State{$r_i = j$}
\EndFor
\end{algorithmic}
\caption{Carpenter-Clifford-Fearnhead sampler}\label{alg:Carpenter}
\end{algorithm}
The CCF sampler conducts on the order of $M/N + 1$ comparisons on average for each $N$ samples, making the average runtime of this algorithm $O(N[M/N + 1]) = O(M + N)$, meaning that this algorithm is linear in both the amount of samples taken and in the size of the distribution form which the samples are taken.
It is important to note that when $N = M$ that this algorithm has a purely linear runtime of $O(N)$, and that when $M \gg N$, this algorithm has a purely linear runtime of $O(M)$, which we improve upon.
This algorithm is not widely known in the particle filtering community as it is notably absent from a highly cited review paper~\cite{li2015resampling}.

\section{Divide and conquer algorithm}
\label{sec:new-algorithm}
The CCF sampler takes advantage of the fact that the both the uniformly distributed samples and the cumulative weights are sorted lists, and performs a simple linear search. As finding the indices of the $U_i$ in the $W_j$ is the exact problem as merging two sorted lists of potentially arbitrarily unequal sizes, we aim to introduce a more sophisticated index finding procedure.

\begin{algorithm}
\begin{algorithmic}
\Require Cumulative weights $\{W_j\}_{j=1}^M$, No. samples $N$
\Ensure $\texttt{DaCSampler}(\{W_j\}_{j=1}^M, N) = \{r_i\}_{i=1}^N$ samples
\State $\{U_i\}_{i=1}^N \gets \texttt{SortedUniform}(N)$
\State $\{r_i\}_{i=1}^{N} \gets \texttt{DaC}(\{U_i\}_{i=1}^{N}, \{W_j\}_{j=1}^{M})$
\Function{\texttt{DaC}}{$\{U_i\}_{i=a_U}^{b_U}$, $\{W_j\}_{j=a_W}^{b_W}$}
\If{$a_U = b_U$}
    \State $r_{a_U} \gets \texttt{BinarySearch}(\{W_j\}_{j=a_W}^{b_W}, U_{a_U})$
\ElsIf{$a_W = b_W$}
    \For{$i = a_U, \dots, b_U$}
    \State $r_i \gets a_W$
    \EndFor
\Else
    \State $m_U \gets \lfloor (a_U + b_U)/2\rfloor$
    \State $m_W \gets \texttt{BinarySearch}(\{W_j\}_{j=a_W}^{b_W}, U_{m_U})$
    \State $r_{m_U} \gets m_W$
    \If{$m_U > a_U$}
    \State $\{r_i\}_{i=a_U}^{m_U - 1} \gets \texttt{DaC}(\{U_i\}_{i=a_U}^{m_U - 1}, \{W_j\}_{j=a_W}^{m_W})$
    \EndIf
    \If{$m_U < b_U$}
    \State $\{r_i\}_{i=m_U + 1}^{b_U} \gets \texttt{DaC}(\{U_i\}_{i=m_U + 1}^{b_U}, \{W_j\}_{j=m_W}^{b_W})$
    \EndIf
\EndIf
\State \Return $\{r\}_{i=a_U}^{b_U}$
\EndFunction
\end{algorithmic}
\caption{Divide and Conquer sampler}\label{alg:Divide-algorithm}
\end{algorithm}

\begin{figure}[ht]
\centering
\tikzset{every picture/.style={line width=1pt}}
\begin{tikzpicture}
\draw[] (-4,0.5) -- (4, 0.5);
\draw[] (-4,0.5) -- (-4, -4.5);
\node (W1T1) at (-3.5, 0) {$W_1$};
\node (W2T1) at (-1.75, 0) {$W_2$};
\node[inner sep=0.0pt] (W3T1) at (0, 0) {$W_3$};
\node (W4T1) at (1.75, 0) {$W_4$};
\node (W5T1) at (3.5, 0) {$W_5$};

\node (U1T1) at (-3,   -0.5) {$U_1$};
\node (U2T1) at (-2.1, -0.5) {$U_2$};
\node[inner sep=0.0pt] (U3T1) at (-1,   -0.5) {$U_3$};
\node[inner sep=0.0pt] (U4T1) at (-0.5, -0.5) {$U_4$};
\node (U5T1) at (3,    -0.5) {$U_5$};
\draw[] (4,0.5) -- (4, -4.5);
\draw[] (-4,-4.5) -- (4, -4.5);
\node at (-4, -4.75) {$0$};
\node at (4, -4.75) {$1$};

\draw[->, dashed] (U3T1) -- (W3T1);

\draw[] (-4,-1) -- (4, -1);

\draw[draw=black, dashed] (-3.8,-0.75) rectangle (3.8,0.25);

\node (W1T2) at (-3.5,  -1.5) {$W_1$};
\node[inner sep=0.0pt] (W2T2) at (-1.75, -1.5) {$W_2$};
\node[inner sep=0.0pt] (W3T2) at (0,     -1.5) {$W_3$};
\node (W4T2) at (1.75,  -1.5) {$W_4$};
\node (W5T2) at (3.5,   -1.5) {$W_5$};

\node[inner sep=0.0pt] (U1T2) at (-3,   -2) {$U_1$};
\node[inner sep=0.0pt] (U2T2) at (-2.1, -2) {$U_2$};
\draw[->, dashed] (U1T2) -- (W2T2);
\draw[draw=black, dashed] (-3.8,-2.25) rectangle (0.3,-1.25);

\node[inner sep=0.0pt] (U4T2) at (-0.5, -2.5) {$U_4$};
\node (U5T2) at (3,    -2.5) {$U_5$};
\draw[->, dashed] (U4T2) -- (W3T2);
\draw[draw=black, dashed] (-0.8,-2.75) rectangle (3.8,-1.25);

\draw[] (-4,-3) -- (4, -3);

\node (W1T3) at (-3.5,  -3.5) {$W_1$};
\node[inner sep=0.0pt] (W2T3) at (-1.75, -3.5) {$W_2$};
\node (W3T3) at (0,     -3.5) {$W_3$};
\node (W4T3) at (1.75,  -3.5) {$W_4$};
\node[inner sep=0.0pt] (W5T3) at (3.5,   -3.5) {$W_5$};

\node[inner sep=0.0pt] (U2T3) at (-2.1,   -4) {$U_2$};
\draw[draw=black, dashed] (-3.8,-4.25) rectangle (-1.45,-3.25);
\node[inner sep=0.0pt] (U5T3) at (3,    -4) {$U_5$};
\draw[draw=black, dashed] (-0.3,-4.25) rectangle (3.8,-3.25);
\draw[->, dashed] (U2T3) -- (W2T3);
\draw[->, dashed] (U5T3) -- (W5T3);

\node [rotate=90,anchor=north] at (-4.5, -0.25) {Step 1}; 
\node [rotate=90,anchor=north] at (-4.5, -2.0) {Step 2}; 
\node [rotate=90,anchor=north] at (-4.5, -3.75) {Step 3}; 

\end{tikzpicture}
\caption{A visual representation of the divide and conquer sampler. The $W$ indicate the cumulative weights from which samples are determined and the $U$ indicate the sorted list of uniform random samples. The dashed lines indicate matching an element to its corresponding weight. Dashed boxes indicate the scope both in $W$ and $U$ of the current steps.}\label{fig:algorithm}
\end{figure}
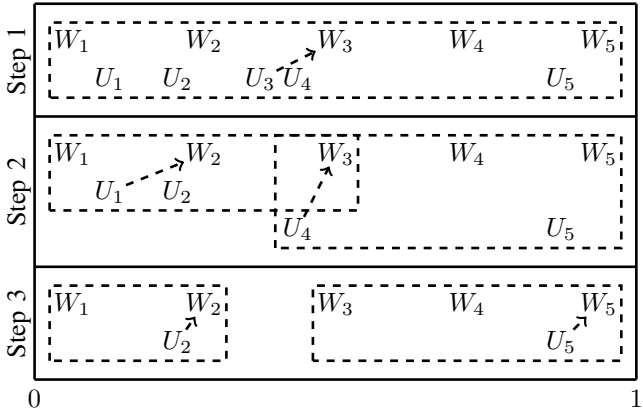

\begin{table}[ht]
    \centering
    \caption{Time complexity of the three multinomial sampling techniques.}
    \label{tab:time-complexity}
    \setlength{\tabcolsep}{3pt}
    %\begin{tabular}{p{100pt}p{100pt}}
    \begin{tabular}{llll}
    \hline
    Algorithm & Time Complexity & $M = N$ & $M \gg N$\\
    \hline
        CCF Sampler &  $O(N[\frac{M}{N}+1])$ & $O(N)$ & $O(M)$\\
        Binary Sampler & $O(N\log_2 M)$ & $O(N\log_2 N)$ & $O(N\log_2 M)$\\
        D\&C Sampler & $O(N\log_2[\frac{M}{N}+1])$ & $O(N)$ & $O(N\log_2[\frac{M}{N}+1])$\\
    \hline\\        
    \end{tabular}
\end{table}

We first discuss two simple base cases that can arise from we build up the rest of the algorithm.
In the case that there is only one cumulative weight $W_j$ that is known and we know that all uniform samples $U_i$ are below this weight, then we know that all samples $s_i$ must correspond to the index $j$, which can be accomplished through a simple loop.
In the opposite case where several cumulative weights $W_j$ are known, but only one uniform sample $U_i$ is known, then the index corresponding to the sample $s_i$ can be found using a binary search.

We are now ready to discuss the case when there are more than one of each $W$ and $U$. As we mean to divide the problem into half, and, as the $U$s are sorted, we find the index of the $U_i$ that corresponds to the middle element. Call this element $U_{m_U}$, and then use binary search to find the weight that corresponds to this element, $W_{m_W}$. The problem can now be split into two, with the first branch consisting of all the $U_i$ with index $i$ less than $m_U$ and all $W_j$ with index $j$ less than or equal to $m_W$, and the second branch consisting of all $U_i$ with index $i$ greater than $m_U$ and all $W_j$ with index $j$ greater than or equal to $m_W$. 
Pseudocode for the algorithm can be found in~\cref{alg:Divide-algorithm}. 
A visual representation of the action of the algorithm for a simple test case can be found in~\cref{fig:algorithm}.

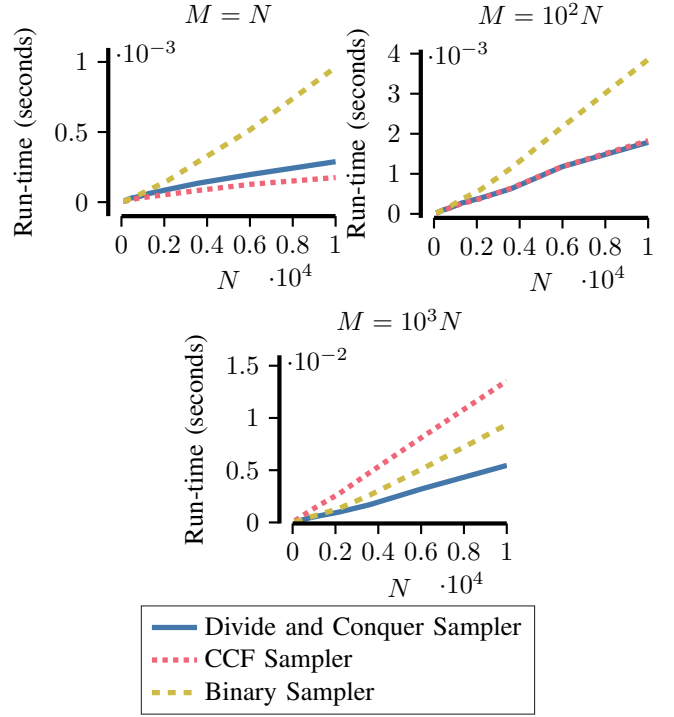
\begin{figure}[ht]
    \centering
    \begin{minipage}{0.5\linewidth}
\begin{tikzpicture}
\begin{axis}[clean,
    width = 1\linewidth,
    %y = 0.33\linewidth,
    cycle list name=tol,
    no markers,
    table/col sep=comma,
    xmin = 0,
    xmax = 10000,
    ymin = -0.0001,
    ymax = 0.0011,
    %clip mode=individual,
    clip = true,
    every y tick scale label/.style={at={(yticklabel cs:0.9,5pt)},xshift=5.5em,yshift=0.5em,left,inner sep=0pt},
    every x tick scale label/.style={at={(xticklabel cs:0.9,5pt)},yshift=0em,left,inner sep=0pt},
    title={$M=N$},
    ylabel = {Run-time (seconds)},
    xlabel = {$N$},
    every axis plot/.append style={line width=2pt, mark size=3.5pt}]

    \addplot[mark=none,color=tolblue] table [x=Ns, y=timesdcM1, col sep=comma] {testresults.csv};
    \addplot[mark=none,dotted,color=tolred] table [x=Ns, y=timesccfM1, col sep=comma] {testresults.csv};
    \addplot[mark=none,dashed,color=tolyellow] table [x=Ns, y=timesbM1, col sep=comma] {testresults.csv};
\end{axis}
\end{tikzpicture}
\end{minipage}%
\begin{minipage}{0.5\linewidth}
\begin{tikzpicture}
\begin{axis}[clean,
    width = 1\linewidth,
    %y = 0.33\linewidth,
    cycle list name=tol,
    no markers,
    table/col sep=comma,
    xmin = 0,
    xmax = 10000,
    ymin = -0.0001,
    ymax = 0.0041,
    %clip mode=individual,
    clip = true,
    every y tick scale label/.style={at={(yticklabel cs:0.9,5pt)},xshift=4.7em,yshift=0.5em,left,inner sep=0pt},
    every x tick scale label/.style={at={(xticklabel cs:0.9,5pt)},yshift=0em,left,inner sep=0pt},
    title={$M=10^2 N$},
    ylabel = {Run-time (seconds)},
    xlabel = {$N$},
    every axis plot/.append style={line width=2pt, mark size=3.5pt}]

    \addplot[mark=none,color=tolblue] table [x=Ns, y=timesdcM100, col sep=comma] {testresults.csv};
    \addplot[mark=none,dotted,color=tolred] table [x=Ns, y=timesccfM100, col sep=comma] {testresults.csv};
    \addplot[mark=none,dashed,color=tolyellow] table [x=Ns, y=timesbM100, col sep=comma] {testresults.csv};
\end{axis}
\end{tikzpicture}
\end{minipage}

\begin{minipage}{0.25\linewidth}
\end{minipage}
\begin{minipage}{0.5\linewidth}
\begin{tikzpicture}
\begin{axis}[clean,
    width = 1\linewidth,
    %y = 0.33\linewidth,
    cycle list name=tol,
    no markers,
    table/col sep=comma,
    xmin = 0,
    xmax = 10000,
    ymin = -0.0001,
    ymax = 0.016,
    %clip mode=individual,
    clip = true,
    every y tick scale label/.style={at={(yticklabel cs:0.9,5pt)},xshift=5.5em,yshift=0.5em,left,inner sep=0pt},
    every x tick scale label/.style={at={(xticklabel cs:0.9,5pt)},yshift=0em,left,inner sep=0pt},
    title={$M=10^3 N$},
    ylabel = {Run-time (seconds)},
    xlabel = {$N$},
    every axis plot/.append style={line width=2pt, mark size=3.5pt}]

    \addplot[mark=none,color=tolblue] table [x=Ns, y=timesdcM1000, col sep=comma] {testresults.csv};
    \addplot[mark=none,dotted,color=tolred] table [x=Ns, y=timesccfM1000, col sep=comma] {testresults.csv};
    \addplot[mark=none,dashed,color=tolyellow] table [x=Ns, y=timesbM1000, col sep=comma] {testresults.csv};
\end{axis}
\end{tikzpicture}
\end{minipage}
\begin{tikzpicture} 
    \begin{axis}[%
    hide axis,
    xmin=10,
    xmax=50,
    ymin=0,
    ymax=0.4,
    cycle list name=tol,
    every axis plot/.append style={line width=2pt, mark size=3.5pt},
    legend style={draw=white!15!black,legend cell align=left, legend columns=1}
    ]
    \addlegendimage{tolblue,mark=none}
    \addlegendentry{Divide and Conquer Sampler};
    \addlegendimage{tolred,dotted,mark=none}
    \addlegendentry{CCF Sampler};
    \addlegendimage{tolyellow,dashed,mark=none}
    \addlegendentry{Binary Sampler};
    \end{axis}
\end{tikzpicture}
\caption{Numerical comparison of the three multinomial sampling algorithms. The $x$ axis in all panels represents the number of samples, $N$, requested, and the $y$ axis represents the runtime in seconds. The three panels represent different sizes of the underlying distribution, $M$.}\label{fig:numerical-experiment}
\end{figure}

Similar to the CCF algorithm in~\cref{alg:Carpenter}, the divide and conquer sampler does roughly $\log_2[M/N + 1]$ operations per each sample $N$, meaning that the time complexity of the divide and conquer sampler is given by $O(N\log_2[M/N + 1])$.
This means that when $M = N$ the time complexity of the divide and conquer sampler is $O(N)$ which means that it behaves similar to the original CCF algorithm and superior to the naive binary search.
When $M \gg N$, the time complexity of the divide and conquer sampler is similar to (but better than) $O(N\log_2(M)$, meaning that in this regime the algorithm behaves similar to (but better than) the naive binary search and superior to the CCF algorithm. This means that the divide and conquer sampler algorithm is the best choice when we know that our problem is either in the $M = N$ or in the $M \gg N$ scenario. A table of time complexities and the two scenarios is given in~\cref{tab:time-complexity}.

\section{Numerical Experiment}
\label{eq:numerical-experiment}

We now numerically show the superior performance of the divide and conquer sampler. 
We test the three algorithms discussed in this work using the ensemble Gaussian mixture filter, which confusingly is a particle filter, specifically the non-adaptive version presented in~\cite{popov2024adaptive}.
The prior collection of particles is taken from the Gaussian distribution defined by the 40-dimensional mean,
\begin{equation}
    [\mu_X]_i = \begin{cases}
        0 & i = 1, \dots 19, 21, \dots 40,\\
        -3.5 & i = 20,
    \end{cases} 
\end{equation}
and the covariance is,
\begin{equation}
    [\Sigma_X]_{i,j} = \begin{cases}
        1 & i = 1, \dots 40, i = j\\
        0.5 & i = 2, \dots 40, j = i - 1\\
        0.5 & i = 1, \dots 39, j = i + 1\\
        0 & \text{sonst}
    \end{cases}.
\end{equation}
A total of $N$ particles is taken is taken from this distribution to generate the prior, and a Gaussian kernel density estimate is constructed using the Silverman bandwidth, resulting in a Gaussian mixture approximation of this prior.

The measurement function is taken to be a range measurement from the origin,
\begin{equation}
    h(x) = \lVert x\rVert_2, 
\end{equation}
with a measurement realization of $y = 1$ with corresponding variance $R = 10^{-2}$. The measurement likelihood is taken to be Laplace distributed,
\begin{equation}
    p_{Y|X} (z) = \frac{1}{\sqrt{2 R}} e^{-\sqrt{\frac{2}{R}}\lvert z - y \rvert},
\end{equation}
and a Gaussian kernel density approximation is constructed,
\begin{equation}
    p_{Y|X} (z) \approx \sum_{i=1}^{N_Y} \mathcal{N}(z ; y_i, \beta_Y^2 \widetilde{R}),
\end{equation}
with a total of $N_Y$ independent samples, using the Silverman bandwidth $\beta_Y^2$~\cite{silverman2018density}, resulting in a Gaussian mixture approximation of the likelihood.

A Gaussian sum update is performed on the prior and likelihood Gaussian mixture models to compute the $M = N\cdot N_Y$ posterior cumulative weights, which are what is important for this work. The posterior cumulative weights are then used to create sample indices using one of the multinomial resampling algorithms described in this work. All timings are computed only on these sampling steps.
For all the experiments our choice of $N$ ranges logarithmically between $10^2$ and $10^4$.
We run three different experiments with $N_Y = 1$, $N_Y = 10^2$, and $N_Y = 10^3$. Timing is measured as a mean for the three algorithms explored in this work over $10^3$ Monte Carlo samples. The code for the numerical experiments is available at \url{https://github.com/AndreyAPopov/Discrete-Sampling}.

The results of these experiments are plotted in figure~\cref{fig:numerical-experiment}.
It can be seen that for the cases $M = N$, meaning that $N_Y = 1$, the divide and conquer sampler is a little more expensive the the CCF sampler, with the binary sampler not competitive with either sampling technique. Both the divide and conquer sampler and the CCF sampler seem linear (with slightly different slope) in $N$, validating the expected computational complexity. For the case $M= 10^2 N$, meaning that $N_Y = 10^2$, the divide and conquer sampler is roughly as expensive as the CCF sampler, with the binary sampler still not competitive. For $M = 10^3 N$, meaning that $N_Y = 10^3$, the CCF sampler is the worst of the three with the binary sampler's $\log_2 M$ term finally winning out over it. The divide and conquer sampler is by far the most efficient of the three in this case, fully validating the computational complexity and the overall approach in general.

\section{Conclusions}
\label{sec:conclusion}
This work provides a new improved resampling algorithm for multinomial sampling for particle filters, with particular emphasis put on sampling for ensemble mixture model filters such as the ensemble Gaussian mixture filter~\cite{anderson1999monte, yun2019sequential, popov2024adaptive}.
We derive the algorithm through a divide and conquer strategy and derive its computational complexity. We show that the algorithm matches both the complexity of the state-of-the-art monomial samplers, namely the CCF sampler and the binary search sampler, and verify this complexity using a numerical experiment.
We give the following recommendation: if the number of samples required is more than or equal to the number of weights, then the CCF algorithm should be used~\cite{carpenter1999improved}, however, if the number of samples required is less than the number of weights, the new divide and conquer sampler presented in this work is the safest choice. 

We end with a note that the techniques outlined in this paper can be generalized to other types of resampling other than monomial such as those presented in~\cite{kuptametee2022review, frisch2021deterministic}, thus purely random resampling is not necessarily required to be performed.

\bibliographystyle{IEEEtran}
\bibliography{biblio}

\end{document}